%% file: paper.tex
\documentclass{ifacconf}
\input{support_files/preamble}

\begin{document}
\begin{frontmatter}
\title{Learning-based cognitive architecture for enhancing coordination in human groups} 

\thanks[footnoteinfo]{This work was in part supported by the Research Project “SHARESPACE” funded by the European Union (EU HORIZON-CL4-2022- HUMAN-01-14.
SHARESPACE. GA 101092889 - http://sharespace.eu).}

\author[SSM]{Antonio Grotta$^\dagger$} 
\author[SSM]{Marco Coraggio$^\dagger$} 
\author[CRdC]{Antonio Spallone}
\author[Unina]{Francesco De Lellis}
\author[SSM,Unina]{Mario di Bernardo$^\ddagger$}

\address[SSM]{Scuola Superiore Meridionale, Naples, Italy.}
\address[CRdC]{CRdC Tecnologie Scarl, Naples, Italy}
\address[Unina]{University of Naples Federico II, Naples, Italy}
\address{$^\dagger$These authors contributed equally}
\address{$^\ddagger$Corresponding author (e-mail: mario.dibernardo@unina.it)}

\begin{abstract}                
As interactions with autonomous agents---ranging from robots in physical settings to avatars in virtual and augmented realities---become more prevalent, developing advanced cognitive architectures is critical for enhancing the dynamics of human-avatar groups. 
This paper presents a reinforcement-learning-based cognitive architecture, trained via a sim-to-real approach, designed to improve synchronization in periodic motor tasks, crucial for applications in group rehabilitation and sports training. 
Extensive numerical validation consistently demonstrates improvements in synchronization. 
Theoretical derivations and numerical investigations are complemented by preliminary experiments with real participants, showing that our avatars can integrate seamlessly into human groups, often being indistinguishable from humans. 
\end{abstract}

\begin{keyword}
Synchronization, Human-avatar groups, Cognitive architecture, Machine learning, Kuramoto model
\end{keyword}

\end{frontmatter}

\section{Introduction}

Human societies are built on group interactions, whether through conversations over shared meals, participation in sports, or involvement in economic and political organizations \citep{homans2017human}. 
As we move towards a future increasingly integrated with virtual and augmented reality, these interactions are poised to
extend to environments where autonomous avatars and robots play significant roles. 
These agents are governed by \emph{cognitive architectures} (CAs) \citep{LANGLEY2009141},
sophisticated control strategies that facilitate their interactions with humans and other avatars. 
The goal of these architectures is to seamlessly integrate avatars into human groups to support specific outcomes, such as enhanced productivity or improved well-being. 
This paper addresses the design of CAs that use feedback control strategies based on reinforcement learning (RL) to enhance coordination and synchronization within human-avatar groups engaged in joint motor tasks. 
Such advancements have promising applications in fields like rehabilitation \citep{howard2017metaanalysis} and sports training \citep{neumann2018systematic} \citep{sharespace}.

Several cognitive architectures have been designed to enhance coordination in motor tasks [describable as combinations of periodic and regulation-to-point dynamics \citep{saltzman1987skilled}]. 
Many examples pertain to the ``mirror game'' \citep{noy2011mirror}, a paradigmatic motor task in which participants are instructed to execute motion that is both ``synchronized and interesting''.
For instance, Zhai et al.~\citep{zhai2014novel} combined the Haken-Kelso-Bunz (HKB) model \citep{haken1985theoretical} and a nonlinear feedback controller to make an avatar play the mirror game, either leading or following the motion of a person.
Subsequently, in \cite{zhai2016design}, a model describing the motion of two players was identified and used to design a CA capable of improvising joint motion with a person. 
A shift towards model-free approaches was then introduced in \cite{lombardi2021using}, employing a Q-learning algorithm to make an avatar interact with a human playing the mirror game in different configurations.

Multiplayer extensions to the mirror game have also been studied, where people in a group perform some joint motor task, e.g., synchronizing the oscillatory motion of their fingers. 
In \cite{alderisio2017interaction}, it was shown that this interaction can be modeled using Kuramoto oscillators. 
Then, in \cite{lombardi2019deep} and \cite{lombardi2021dynamic}, CAs based on RL were presented, with the goal of displaying a specific \emph{motor signature}---a quantity shown to uniquely identify different people \citep{slowinski2016dynamic}. 

In this study, we address the design of cognitive architectures to enhance synchronization—linked to participants' well-being \citep{rennung2016prosocial}—in groups of humans and avatars performing periodic motor tasks, which are relevant for rehabilitation \citep{howard2017metaanalysis} and sports training \citep{neumann2018systematic}. 
We provide an analytical result to guide the design for groups of minimum size; then to the best of our knowledge, we introduce the first reinforcement-learning-based CA to drive an avatar to improve synchronization and coordination in a group of humans. 
The CA we propose is trained using synthetic data from Kuramoto networks. 
Extensive numerical validation demonstrates consistent synchronization improvements when the CA is used. 
Preliminary experiments with real people, 
show the avatar integrates effectively with humans, maintaining coordination and often being indistinguishable from a human participant.

\section{Problem description}%
\label{sec:problem_description}

In what follows we denote by 
$\BB{S} = \BB{R} / 2\pi \BB{Z}$ the 1-sphere.
$\R{sat}_{v_\R{min}}^{v_\R{max}}(v) \coloneqq \R{min}(\R{max}(v, v_\R{min}) v_\R{max})$ indicates a saturation.
In a vector, $\parallel$ denotes concatenation, while $\imag \coloneqq \sqrt{-1}$.

As in \cite{alderisio2017interaction}, we model the motion of a group of $n$ humans and avatars performing a periodic motor task akin to a multiplayer mirror game \citep{noy2011mirror} as a network of Kuramoto oscillators of the form
\begin{equation}\label{eq:kuramoto_model}
    \frac{\R{d}\theta_i(t)}{\R{d}t} = \theta_i(t) + 
        \omega_i(t) + c \sum_{j = 1}^n A_{i j} \sin \left(\theta_j(t)-\theta_i(t) \right),
\end{equation}
for $i \in \{1, \dots, n\}$, where
$t \in \BB{R}$ is continuous time, 
$\theta_i(t) \in \BB{S}$ is the phase of participant $i$, 
$\omega_i(t) \in \BB{R}_{> 0}$ is their \emph{natural frequency}, 
$c \in \BB{R}_{\ge 0}$ is a \emph{coupling gain} 
modeling the strength of the interaction among participants, 
and $[A_{ij}] \in \{0, 1\}^{n \times n}$ is the \emph{adjacency matrix} describing the interaction pattern in the group 
($A_{ii} = 0 \ \forall i$; $A_{ij}$ being $1$ if node $j$ influences $i$, or $0$ otherwise).

Let $\{\C{I}_\R{p}, \C{I}_\R{a}\}$ be a partition of $\{1, \dots, n \}$, with indices in $\C{I}_\R{p}$ representing (real or simulated) human participants, and those in $\C{I}_\R{a}$ representing the autonomous avatars;
let also $n_\R{p} \coloneqq \abs{\C{I}_\R{p}}$, $n_\R{a} \coloneqq \abs{\C{I}_\R{a}}$.

Denote $\theta_{i,j}(t) \coloneqq \theta_i(t) - \theta_j(t)$;
let a \emph{configuration} be some
\begin{equation*}
    \phi \coloneqq [
    \theta_{1,2} \ \theta_{2,3} \ \cdots \ \theta_{n-1,n}]\T \in \BB{S}^{n-1}.
\end{equation*}
Some $\bar{\phi} \in \BB{S}^{n-1}$ is a \emph{phase-locked configuration} if $\dot{\phi}(\bar{\phi}) = 0$, where $\dot{\phi}$ is obtained from \eqref{eq:kuramoto_model}.

We assess the level of phase synchronization in the network via the \emph{order parameter}. Specifically, let 
$q_{\R{tot}}(t) \coloneqq \frac{1}{n} \sum_{i=1}^n e^{\imag \theta_i(t)}$
be the \emph{total average phasor} and 
$q_{\R{net}}(t) \coloneqq \frac{1}{n_\R{p}} \sum_{i \in \C{I}_\R{p}} e^{\imag \theta_i(t)}$
be the \emph{net average phasor}.
The \emph{total order parameter} is $r_\R{tot}(t) \coloneqq \abs{q_\R{tot}(t)}$ and is computed considering all participants in the group including the avatars;
the \emph{net order parameter} is $r_\R{net}(t) \coloneqq \abs{q_\R{net}(t)}$ and is computed instead considering all participants {\em but} the avatars.
Note that $r_\R{tot}(t), r_\R{net}(t) \in [0, 1]$; values equal to $0$ and $1$ being associated to states of low and high phase synchronization, respectively.
Then, define the time-average
$\langle r_\R{net} \rangle \coloneqq \tfrac{1}{T} \int_{0}^T r_\R{net}(\tau) \, \R{d}\tau$, where $T \in \BB{R}_{>0}$ is the length of the interaction, and
$\langle r_\R{tot} \rangle \coloneqq \tfrac{1}{T} \int_{0}^T r_\R{tot}(\tau) \, \R{d}\tau$.

We aim to control the avatar dynamics by modulating the frequencies $\omega_i$, $i \in \C{I}_\R{a}$, to solve the optimization problem
\begin{equation}\label{eq:optimization_problem}
    \begin{aligned}
        \max\nolimits_{\{\omega_i\}_{i \in \C{I}_\R{a}}}\ \ &\, \langle r_\R{net} \rangle,
    \end{aligned}
\end{equation}
corresponding to maximizing coordination among human participants.
Below, we assume the presence of one avatar, ($n_\R{a} = 1$) and $n-1$ human participants. 
W.l.o.g., we will refer to the avatar as the $n$-th agent, labeling its phase and frequency with the subscript $\R{a}$ or $n$. 

\section{Analysis}

As a first step, we study the simplest scenario consisting of $n = 3$ participants (2 humans and 1 avatar). Assuming $\omega_1,\omega_2$ are constant, and that participants are connected in a complete graph (i.e., $A_{ij} = 1 \ \forall i,j$), we wish to assess if and how the avatar can modulate its dynamics to maximize coordination among the human participants. 

To this aim, given some $c$ and $\{\omega_i\}_{i \in \C{I}_\R{p}}$, let $\Phi \subset \BB{S}^{n-1}$ denote the set of all possible phase-locked configurations obtainable by varying $\omega_\R{a} \in \BB{R}_{\ge 0}$.

\begin{thm}\label{thm:best_phase_lock_configuration}
    Assume
    $\C{I}_\R{p} = \{1, 2\}$, 
    $\C{I}_\R{a} = \{3\}$.
    Let
    $\chi \coloneqq 2\cos^{-1}\tfrac{-1 + \sqrt{33}}{8}$ and 
    $\nu \coloneqq \sin \chi + \sin \tfrac{\chi}{2}$.
    If $\abs{\tfrac{\omega_1 - \omega_2}{2c}} < \nu$
    and $\omega_\R{a} = (\omega_1 + \omega_2) / 2$, then (i) there exists a phase-lock configuration $\bar{\phi}^\star \in (\chi, \chi) \times \{ \tfrac{\theta_{1,2}}{2}\}$, and   
    (ii) $\bar{\phi}^\star = \arg \max_{\phi \in \Phi} r_\R{net}$.
\end{thm}

\begin{pf}
Define $\theta_\R{m} \coloneqq \ON{arg}\left(e^{\imag\theta_1} \right. + \left. e^{\imag\theta_2}\right)$ as the average phase between $\theta_1$ and $\theta_2$, and let $\epsilon \coloneqq \theta_\R{a} - \theta_\R{m}$, so that 
\begin{equation}\label{eq:transformation_differences}
    \theta_{\R{a},1} = - \tfrac{\theta_{1,2}}{2} + \epsilon;
    \quad
    \theta_{\R{a},2} =   \tfrac{\theta_{1,2}}{2} + \epsilon.    
\end{equation}
A configuration $\bar{\phi} \coloneqq [\bar{\theta}_{1,2} \ \bar{\theta}_{2,\R{a}}]\T \in \BB{S}^2$ (with associated $\bar{\epsilon}$) is a phase-lock one if 
$\dot{\phi}(\bar{\phi}) = 0$.
Hence, from \eqref{eq:kuramoto_model}, we get
\begin{equation}\label{eq:dynamics_configuration}
\begin{dcases}
    \dot{\theta}_{1,2} = 
    \omega_1 - \omega_2 + c ( 
    \sin \theta_{2,1} +
    \sin \theta_{\R{a},1} -
    \sin \theta_{1,2} -
    \sin \theta_{\R{a},2}),
    \\
    \dot{\theta}_{1,\R{a}} = 
    \omega_2 - \omega_\R{a} + c ( 
    \sin \theta_{2,1} +
    \sin \theta_{\R{a},1} -
    \sin \theta_{1,\R{a}} -
    \sin \theta_{2,\R{a}}).
    \end{dcases}
\end{equation}
Imposing $\dot{\theta}_{1,2} = \dot{\theta}_{1,\R{a}} = 0$,
using \eqref{eq:transformation_differences} and trigonometric identities ($\sin(\alpha \pm \beta) = \sin \alpha \cos \beta \pm \cos\alpha \sin\beta$), we rewrite
\begin{align}
    2 \sin \bar{\theta}_{1,2} + 2 \sin \tfrac{\bar{\theta}_{1,2}}{2}  \cos \bar{\epsilon} &= \tfrac{\omega_1 - \omega_2}{c},\label{eq:phase_lock_12}\\
    \sin \bar{\theta}_{1,2} + \sin \left( \tfrac{\bar{\theta}_{1,2}}{2} + \bar{\epsilon} \right) - 2 \sin{\bar{\epsilon}}\cos{\tfrac{\bar{\theta}_{1,2}}{2}} &= \tfrac{\omega_1 - \omega_\R{a}}{c}.\label{eq:phase_lock_13}
\end{align}
We will now seek phase-locked configurations that minimize $\abs{\theta_{1,2}}$, as this in turn maximizes $r_\R{net}$.
From \eqref{eq:phase_lock_12}, it can be easily verified by plotting the graphs of the functions involved that $\abs{\bar{\theta}_{1,2}}$ is minimized when $\bar{\epsilon} = 0$. Hence, we set $\bar{\epsilon} = 0$.
Moreover, as $\cos \bar{\theta}_{1,2} = 2 \cos^2\tfrac{\bar{\theta}_{1,2}}{2}-1$, we find that, for $\bar{\theta}_{1,2} \in [-\pi, \pi]$, the function $\sin \bar{\theta}_{1,2} + \sin \tfrac{\bar{\theta}_{1,2}}{2}$ has extrema in $\pm \chi$, being equal to $\pm \nu$ ($\nu$ and $\chi$ defined in the theorem statement).
Hence, if $\abs{\tfrac{\omega_1 - \omega_2}{2c}} \in [-\nu, \nu]$, \eqref{eq:phase_lock_12} has (at least) a solution in $\bar{\theta}_{1,2}^\star \in [-\chi, \chi]$.
Moreover, if $\omega_\R{a} = (\omega_1 + \omega_2) / 2$, then $\bar{\theta}_{1,2}^\star$ is also a solution to \eqref{eq:phase_lock_13}, confirming $\bar{\phi}^\star \coloneqq [\bar{\theta}_{1,2}^\star \ \tfrac{\bar{\theta}^\star_{1,2}}{2}]\T$ is phase-locked.~\hfill$\qed$
\end{pf}

Additionally, local asymptotic stability of $\bar{\phi}^\star$ in Theorem \ref{thm:best_phase_lock_configuration} can be assessed by evaluating the eigenvalues of the Jacobian matrix $J$ of the ODE in \eqref{eq:dynamics_configuration}, given by
\begin{equation*}
    J = c\left[\begin{smallmatrix}
        -2\cos\theta_{1,2} - \cos(-\theta_{1,\R{a}} + \theta_{1,2}) &
        -\cos \theta_{1,\R{a}} + \cos(-\theta_{1,\R{a}} + \theta_{1,2}) \\
        -\cos\theta_{1,2} + \cos(-\theta_{1,2} + \theta_{1,\R{a}}) &
        -2\cos\theta_{1,\R{a}} - \cos(-\theta_{1,2} + \theta_{1,\R{a}})
    \end{smallmatrix}\right].
\end{equation*}
Numerical inspection shows that both eigenvalues of $J$ are negative for $\left( \theta_{1,2}, \theta_{1,\R{a}} \right) \in (-\chi, \chi) \times \lbrace \tfrac{\theta_{1,2}}{2} \rbrace$, ensuring local asymptotic stability of $\bar{\phi}^\star$.

Theorem \ref{thm:best_phase_lock_configuration} suggests that selecting $\omega_\R{a} = \tfrac{\omega_1 + \omega_2}{2}$ yields the phase-locked configuration with the largest $r_\R{net}$.
However, the analysis is more complex for $n > 3$, and hence we resort to a reinforcement learning approach to decide $\omega_\R{a}$.

\section{Design of the cognitive architecture}%
\label{sec:control_solution}

\IncMargin{0.9em}
\begin{algorithm2e}[t]
  \caption{Phase and amplitude estimation}
  \label{alg:phase_estimation}
  \small
  \KwIn{%
  Positions $p_{t-1}$, $p_t$;
  velocities $v_{t-1}$, $v_{t}$
  amplitudes $A^{p>0}_{t-1}$, $A^{p<0}_{t-1}$, $A^{v>0}_{t-1}$, $A^{v<0}_{t-1}$.
  }
  \KwOut{%
  phase $\theta_t$; 
  amplitudes $A^{p > 0}_{t}$, $A^{p < 0}_{t}$, $A^{v>0}_{t}$, $A^{v<0}_{t}$.
  }
  \BlankLine
    \leIf{$p_{t-1} < 0 \wedge p_t \ge 0$,}
        {$A^{v>0}_t = \abs{v_t}$,}
        {$A^{v>0}_t = A^{v>0}_{t-1}$}
    \leIf{$p_{t-1} \ge 0 \wedge p_t < 0$,}
        {$A^{v < 0}_t = \abs{v_t}$,}
        {$A^{v < 0}_t = A^{v < 0}_{t-1}$}
    \leIf{$v_{t-1} \ge 0 \wedge v_t < 0$,}
        {$A^{p>0}_t = \abs{p_t}$,}
        {$A^{p>0}_t = A^{p>0}_{t-1}$}
    \leIf{$v_{t-1} < 0 \wedge v_t \ge 0$,}
        {$A^{p < 0}_t = \abs{p_t}$,}
        {$A^{p < 0}_t = A^{p < 0}_{t-1}$}
    \BlankLine
    \leIf{$p_{t} \ge 0$,}
        {$p_t^{\R{norm}} = p_t / A^{p>0}_t$,}
        {$p_t^{\R{norm}} = p_t / A^{p < 0}_t$}
    \leIf{$v_{t} \ge 0$,}
        {$v_t^{\R{norm}} = v_t / A^{v>0}_t$,}
        {$v_t^{\R{norm}} = v_t / A^{v < 0}_t$}
    \BlankLine
    $\theta_t = \R{atan2}( -v_t^{\R{norm}}, p_t^{\R{norm}})$\;
\end{algorithm2e}
\DecMargin{0.9em}

\begin{figure}[t]
   \centering
   \includegraphics[max width=\columnwidth]{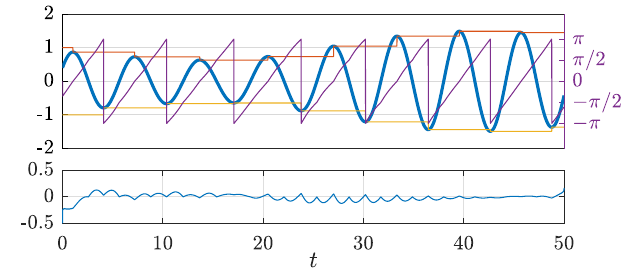}
   \caption{(Top) Example of phase reconstruction (purple) from a position signal (blue) through Algorithm \ref{alg:phase_estimation}; orange and gold lines are $A^{p > 0}(t)$ and $A^{p < 0}(t)$.
   (Bottom) Difference between phases estimated with the Hilbert transform (offline) and Algorithm \ref{alg:phase_estimation} (online).}
   \label{fig:estimation_phase_amplitude}
\end{figure}

To design the CA,
we discretize \eqref{eq:kuramoto_model} via a forward Euler scheme, with sampling time $\Delta t \in \BB{R}$, and denote discrete time by $k \in \BB{Z}$.
Then, we adapt $\omega_\R{a}$ online according to
\begin{equation}
    \omega_\R{a} (k+1) = \R{sat}_{\omega_\R{min}}^{\omega_\R{max}}(\omega_\R{a} (k) + \Delta\omega(k)),
\end{equation}
where $\Delta\omega(k)$ is decided by a RL strategy and $\omega_\R{min}, \omega_\R{max} \in \BB{R}$.
The state of the RL model is
$\xi(k) = [\arg q_\R{a}(k) \parallel 1-\abs{q_\R{a}(k)} \parallel \omega_\R{a}(k)]\T$, 
where
$q_\R{a} \coloneqq \frac{1}{n_\R{p}} \sum_{i \in \C{I}_\R{p}} e^{\imag \left(\theta_{\R{a}}-\theta_i\right)}$;
$\arg q_\R{a}$ is the average phase of the group w.r.t~the avatar, while $1-\abs{q_\R{a}}$ is a variance \citep{fisher1993statistical}.
Hence, the state space is $\BB{S} \times [0, 1] \times [\omega_\R{min}, \omega_\R{max}]$;
the control input is $\Delta\omega(k) \in \{-0.5, -0.4, \dots, 0.5\}$;
the reward is chosen as $r_\R{tot}^2(k)$.%
\footnote{Although Problem \eqref{eq:optimization_problem} involves $r_\R{net}$ rather than $r_\R{tot}$, we verified numerically that using the latter as a reward yields significantly better performance.}
%

\subsection{Training of the cognitive architecture}

To implement reinforcement learning, we utilized a Deep Q-Network (DQN) approach \citep{mnih2015humanlevel}.
Our neural network architecture (approximating the state-action value function $Q$) was chosen heuristically and includes $3$ nodes in the input layer, $2$ hidden layers with $128$ and $64$ nodes respectively, with ReLU activation functions, and an output layer of $11$ nodes with linear activation. 
During training, we employed an $\varepsilon$-greedy policy with $\varepsilon=0.1$, using an Adam optimizer with a learning rate of $0.001$. 
The discount factor was set to $0.9$, with a replay buffer capacity of $100\,000$ and a batch size of $32$. Mean square error served as the loss function.

Given the impracticality of training the RL model with real participants, synthetic data was generated using the Kuramoto network model \eqref{eq:kuramoto_model}.
We set $\Delta t = 0.01\,\text{s}$, $c = 1.25$, $n_\R{p} = 2$, and utilized a complete interaction graph.
Training involved $500$ episodes, each consisting of a $5\,\text{s}$ simulation. 
Natural frequencies $\{\omega_i\}_{i \in \C{I}_p}$ were uniformly drawn from $[4-\delta_\omega, 4+\delta_\omega]$ with $\delta_\omega=0.6$, and initial phases randomly selected from $[\pi/4, 3\pi / 4]$ to facilitate phase locking [based on what reported in \cite{alderisio2017interaction} and \cite[Thm.~17.9]{bullo2022lectures}].


\section{Numerical validation}
\label{sec:simulations}

\begin{figure}[t]
   \centering
   \subfloat[]{\includegraphics[max height=4cm]{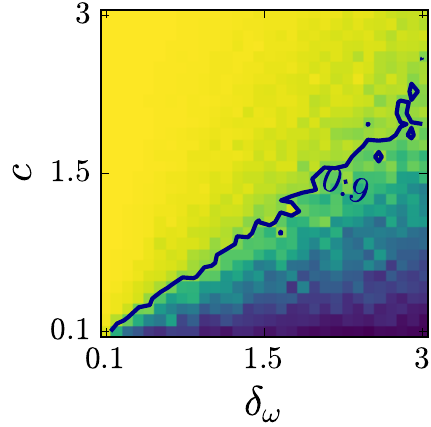}
    \label{fig:robustness_without_ca}}
    \hfill
   \subfloat[]{\includegraphics[max height=4cm]{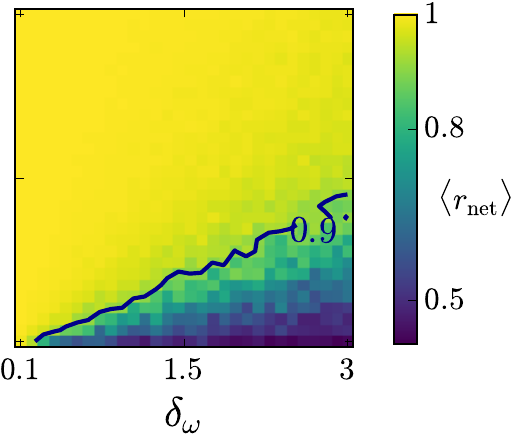}
    \label{fig:robustness_with_ca}}
   \caption{%
  Enhancement of synchronization achieved by the CA.
  $n_\R{p}=5$, connected in a ring.
  (a) without CA ($n_\R{a} = 0$, $n = 5$); (b) with CA connected to all ($n_\R{a} = 1$, $n = 6$).
  Each pixel is the average of $15$ simulations: in each, 
  $\{\omega_i\}_{i \in \C{I}_\R{p}}$ drawn randomly in $[4 - \delta_{\omega}, 4 + \delta_{\omega}]$. 
  $\theta_i(k=0) = \pi/2, \forall i$.
  $T = 100 \,\text{s}$.
  }
   \label{fig:robustness}
\end{figure}

\begin{table}[t]
  \centering
  \begin{tabular}{l|l|llll}
      \toprule
      Reference & $n_\R{p}$ &  \multicolumn{4}{c}{\% increase}\\{} & {} & $\langle r_\R{net} \rangle$  &  $\langle r_\R{tot} \rangle$  &  $\langle \rho_\R{net} \rangle$  &  $\langle \rho_\R{tot} \rangle$\\
      \midrule
      Ref.~1, group 1 & 7  &  63.6\% &  68.7\%  &  49.8\%  &  52.4\% \\
      Ref.~1, group 2 & 7 &  57.5\% &  58.0\%  &  19.6\%  &  19.6\% \\
      Ref.~2 & 5 & 17.8\% & 20\%  &  10.4\%  &  10.4\% \\
      \bottomrule
  \end{tabular}
  \vspace{0.5em}
  \caption{%
  Enhancement of synchronization achieved by the CA.
  Participants connected in a ring; CA connected to all.
  $c$ and $\{\omega_i(k)\}_{i \in \C{I}_\R{p}}$ (time-varying, drawn from Gaussian distributions) taken from the literature: Ref.~1 is \cite{alderisio2017interaction}, Ref.~2 is \cite{calabrese2022modeling}.
  $\theta_i(k=0) = \pi/2, \forall i$. $T = 300\,\text{s}$. 
  Each simulation is repeated 5 times.
  $\langle \rho_\R{net} \rangle$,  $\langle \rho_\R{tot} \rangle$ measure frequency synchronization, as defined in the Appendix.
  }
  \label{tab:improvement_parameters_literature}
\end{table}


To validate the CA presented in \S~\ref{sec:control_solution}, in Figure \ref{fig:robustness}, we report the value of $\langle r_\R{net} \rangle$ 
obtained in a group
in the absence (Fig.~\ref{fig:robustness_without_ca})
and in the presence (Fig.~\ref{fig:robustness_with_ca}) of an avatar driven by the CA, under different values of coupling gains and group frequencies.
Figure \ref{fig:robustness} shows a significant increase of the region where $\langle r_\R{net} \rangle \ge 0.9$ (yellow area) when the avatar is present, confirming the effectiveness of the CA in adapting the avatar dynamics to improve the synchronization level of the other participants.
Additional validation is reported in Table \ref{tab:improvement_parameters_literature}, showing the performance improvement achieved by adding an avatar driven by our CA to groups modeled as Kuramoto oscillators parameterized from real data available in the literature.

\begin{figure}[t]
    \centering
    \includegraphics[max width=\columnwidth]{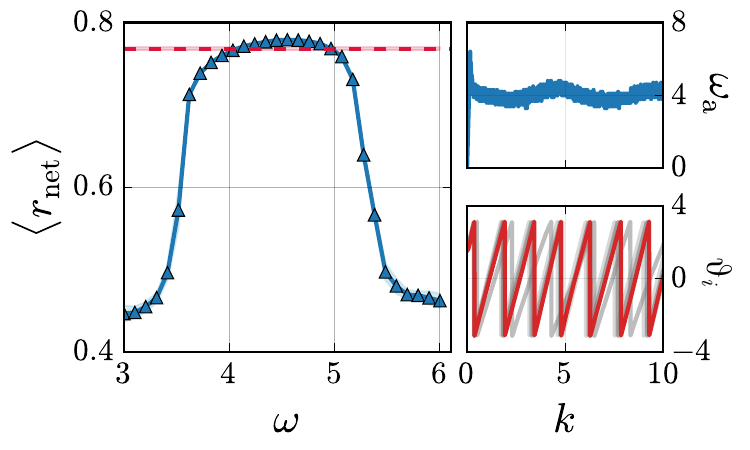}
    \caption{%
    Optimality of the behavior of the CA.
    $n_\R{p} = 7$, connected in a ring.
    $c$ and $\{\omega_i(k)\}_{i \in \C{I}_\R{p}}$ (time-varying, drawn from Gaussian distributions) taken from \cite[group 1]{alderisio2017interaction}.
    $\theta_i(k=0) = \pi/2, \forall i$. $T = 300\,\text{s}$.
    (Left) The blue line and shading are mean and standard deviation (simulations repeated $5$ times) obtained by an agent having frequency $\omega$ ($x$-axis) added to the group and connected to all;
    red dashed line and shading are obtained by a CA added and connected to all.
    (Top right) $\omega_\R{a}(k)$ selected by the CA.    
    (Bottom right) $\{\theta_i(k)\}_{i \in \C{I}_\R{p}}$ (gray) and $\theta_\R{a}(k)$ (red) in the first (of $5$) simulation.
   }
   \label{fig:bell_curve}
\end{figure}

Next, we verify that the CA is able to select the frequency $\omega_\R{a}$ that improves $\langle r_\R{net} \rangle$ the most. 
In Figure \ref{fig:bell_curve}, we consider a group parametrized from real data, by having $\{\omega_i(k)\}_{i \in \C{I}_\R{p}}$ drawn at random (at each time step $k$) from $n_\R{p}$ Gaussian distributions with means and variances taken from \cite[group 1]{alderisio2017interaction}, from which we also take the value of $c$.
Then, we compare the value of $\langle r_\R{net} \rangle$ obtained by adding a CA (red dashed line) against that obtained by adding a non-adaptive agent with fixed frequency instead.
Indeed,  the CA achieves performance close to the optimum (see the left panel) by selecting frequencies that oscillate around $4\,\text{rad/s}$ (see the top right panel), closely matching the participants' average of $4.15\,\text{rad/s}$, in accordance with the insight from Theorem \ref{thm:best_phase_lock_configuration} (holding for $n=3$).

\begin{figure}[t]
    \centering
    \includegraphics[max width=\columnwidth]{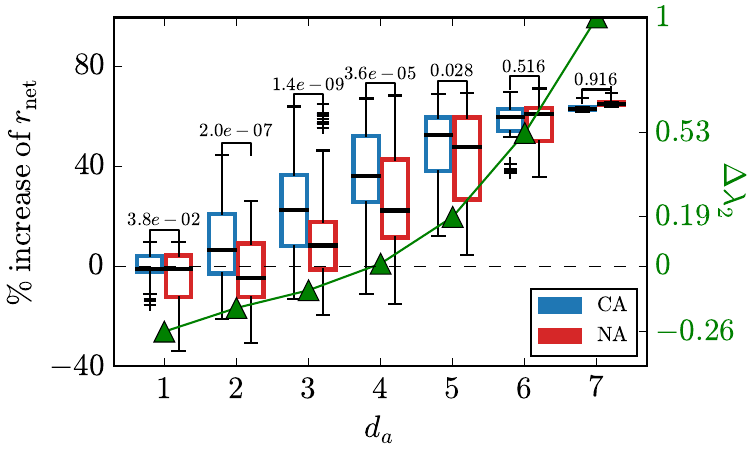}
    \caption{%
    Effect of node degree on synchronization performance and comparison between the CA and a naive avatar.
    $n_\R{p} = 7$, connected in a ring.
    $c$ and $\{\omega_i(k)\}_{i \in \C{I}_\R{p}}$ (time-varying, drawn from Gaussian distributions) taken from \cite[group 1]{alderisio2017interaction}.
    $\theta_i(k=0) = \pi/2, \forall i$. $T = 100\,\text{s}$.
    Box plots are distributions over all possible arrangements of the CA/NA's edges, and $5$ simulations per case.    
    Numbers in black are $p$-values of the null hypothesis that the mean performance of CA is not superior to that of NA (one-tailed $t$-tests).
    $\Delta \lambda_2 \coloneqq \lambda_2^{\R{a}} - \lambda_2^{\neg\R{a}}$, where $\lambda_2^{\R{a}}$ and $\lambda_2^{\neg\R{a}}$ are the algebraic connectivity when the avatar (CA/NA) is absent and present, respectively.    
    }
    \label{fig:box_plot}
\end{figure}


In Figure \ref{fig:box_plot}, we consider again a group parameterized using real data from \cite[group 1]{alderisio2017interaction}, as in the previous simulation study. We evaluate the performance of the CA when it is connected to different numbers of participants---i.e., varying its \emph{node degree} $d_\R{a}$---and compare it to a \emph{naive agent} (NA), which sets $\omega_\R{a}(k) = f \left( \tfrac{1}{n_\R{p}} \sum_{i = 1}^{n_\R{p}} \dot\theta_i(k) \right)$ where $f$ is a discrete-time filter with unitary gain and a pole at $30\,\text{Hz}$.
Figure \ref{fig:box_plot} shows that, on average, the larger the degree $d_\R{a}$ of the artificial agent (either driven by a CA or being a NA), the higher the improvement in $\langle r_\R{net} \rangle$ is.
Moreover, a one-tailed $t$-test shows that the CA outperforms the NA for $d_\R{a} \in \{1, \dots, 5\}$, whereas the performance of the two are comparable for $d_\R{a} \in \{6, 7\}$.
Note that the decrease in $\langle r_\R{net} \rangle$ obtained by both the CA and the NA for low values of $d_\R{a}$ is likely mostly due to the communication graph becoming more difficult to synchronize, as shown by the decrease in the algebraic connectivity (green line).

\section{Preliminary experimental validation}
\label{sec:experiments}

\subsection{Description of experimental setup}
\label{sec:experiments_description}
\begin{figure}[t]
   \centering
   \subfloat[]{\includegraphics[max width=0.49\columnwidth]{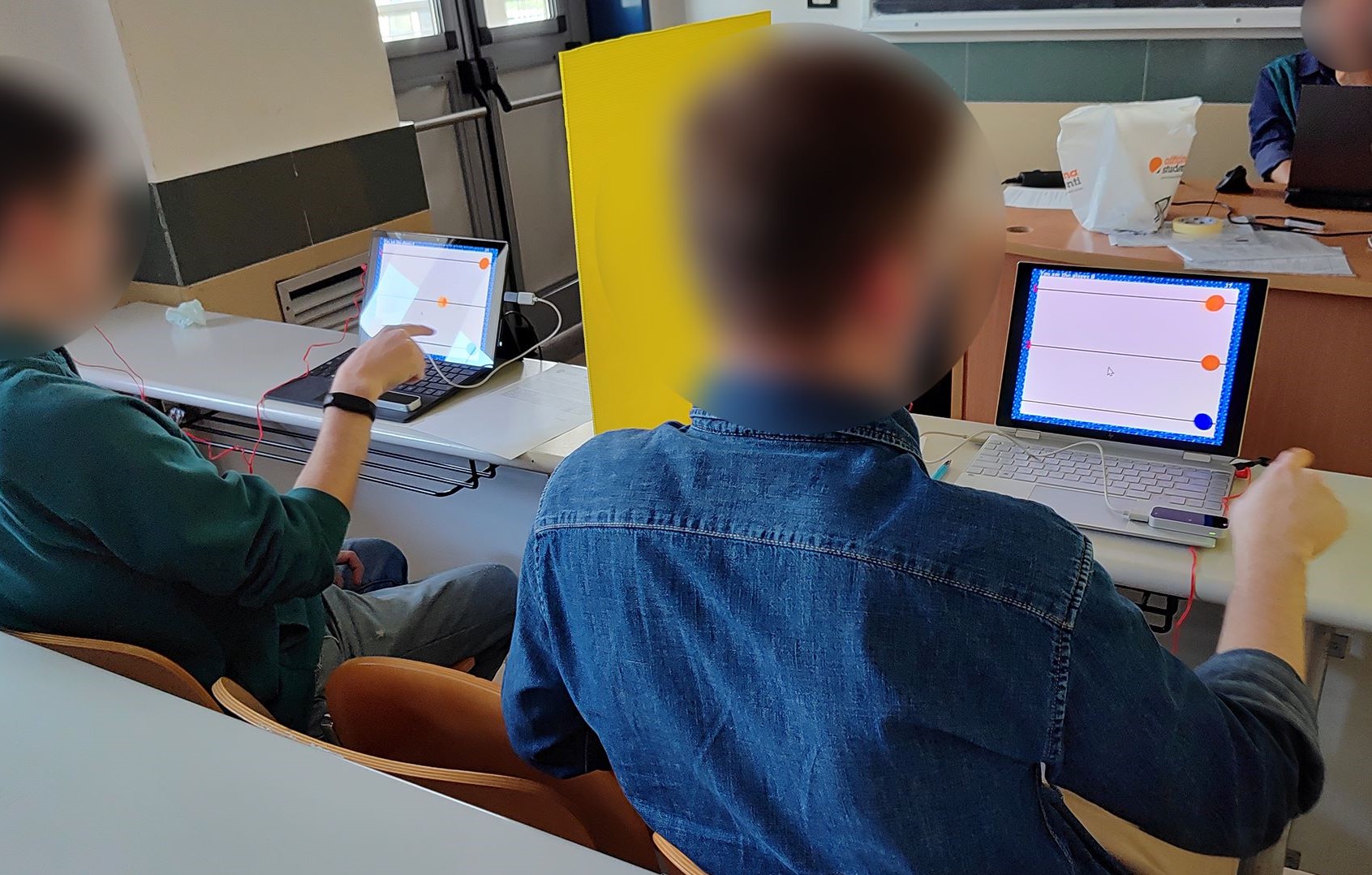}
    \label{fig:experiment_photo}}
    \hfill
   \subfloat[]{\includegraphics[max width=0.4\columnwidth]{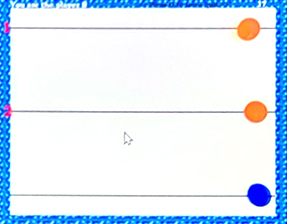}
       \label{fig:experiment_screen}}
   \caption{(a) Experimental setup.
   (b) Chronos interface.}
   \label{fig:experiment}
\end{figure}

To further validate our cognitive architecture, we conducted experiments with three groups of five human participants each, utilizing ``Chronos''%
\footnote{\url{https://dibernardogroup.github.io/Chronos/download.html}}
\citep{alderisio2017novel}, a software we extended for this research. Participants were placed in a room, visually and acoustically isolated from each other, each seated at a desk equipped with a ``Leap Motion Controller''---a small hand tracking device using infrared cameras---and a computer running Chronos. 
The position of each participant's hand was captured at $40\,\text{Hz}$ (Fig.~\ref{fig:experiment_photo}) and displayed on the computer screen as a blue ball moving horizontally (Fig.~\ref{fig:experiment_screen}). 
The screen also displayed orange balls, representing other participants' motion according to a predetermined interaction structure.
Participants were instructed to move in a smooth oscillatory fashion, from left to right and back, synchronizing their motion with that of others. They were informed that in some trials, an autonomous agent might control one of the orange balls, though its activation was not disclosed.
Participants' phases were computed online from the balls' positions on the screen as explained in Section \ref{sec:phase_position}.

We designed six experimental conditions, each lasting $30\,\text{s}$:
\begin{itemize}
\item
\emph{Solo}: Participants do not interact.
\item
\emph{People only (P)}: Participants interact according to a ring graph.
\item
\emph{Cognitive architecture (CA)}: Similar to P, but with a CA visible to all participants (cf.~\S~\ref{sec:control_solution}).
\item
\emph{Naive avatar (NA)}: Similar to P, but with a naive avatar visible to all (cf.~\S~\ref{sec:simulations}).
\item
\emph{CA replacing closest (CA-RC)}: Similar to P, but a CA replaces the participant whose frequency is closest to the group's mean.
\item
\emph{CA replacing farthest (CA-RF)}: Similar to P, but the CA replaces the participant farthest from the mean.
\end{itemize}

Initially, we ran the Solo condition five times to estimate the natural frequencies ($\omega_i$) of the participants, informing conditions CA-RC and CA-RF. Then, we ran conditions P, CA, NA, CA-RC, and CA-RF, repeating each across five trials and mixing all trials randomly.
After each trial, participants were asked if they could identify whether they had interacted with an avatar and, if so, which was its ball.

\subsection{Reconstruction of phase and position from each other}
\label{sec:phase_position}

Most existing methods to extract a phase from the position signal of a participant's hand [e.g., Hilbert transform \cite[see Sec. A2.1]{pikovskij2003synchronization}] require long portions of the signal and cannot be applied online.
Hence, we use the ad-hoc Algorithm \ref{alg:phase_estimation}, inspired by \cite{mortl2012modeling}.
The Algorithm assumes that the signal oscillates around $0$,%
\footnote{Formally, $\forall t_1, t_2$ s.t.~$t_1 \ne t_2$ and $p(t_1) = p(t_2) = 0$, $\exists t_3 \in [t_1, t_2]$ s.t.~$t_3 \ne t_1$, $t_3 \ne t_2$ and $v(t_3) = 0$;
moreover, 
$\forall t_1, t_2$ s.t.~$t_1 \ne t_2$ and $v(t_1) = v(t_2) = 0$, $\exists t_3 \in [t_1, t_2]$ s.t.~$t_3 \ne t_1$, $t_3 \ne t_2$ and $p(t_3) = 0$. Additionally, $\nexists t$ s.t.~$p(t)=v(t)=0$.}
possibly with time-varying frequency and amplitude;
it returns the current phase $\theta(t)$ of signal $p$ with velocity $v$, and $A^{p > 0}(t)$, $A^{p < 0}(t)$, being the current amplitudes when $p > 0$ and $p < 0$, respectively.
A graphical example of the estimation procedure is reported in Figure \ref{fig:estimation_phase_amplitude}.

To transform the avatar's phase $\theta_\R{a}(t)$ into a position $p(t)$, representable as a ball on computer screens, we computed
\begin{equation*}
        p(t) = \begin{dcases}
            A_\R{a}^{p > 0}(t) \cos \theta_\R{a}(t), &\theta_\R{a}(t) \in [-\tfrac{\pi}{2},  \tfrac{\pi}{2}],\\
            A_\R{a}^{p < 0}(t) \cos \theta_\R{a}(t), &\theta_\R{a}(t) \in [-\pi, -\tfrac{\pi}{2}) \cup (\tfrac{\pi}{2}, \pi].
        \end{dcases}
\end{equation*}
where 
$A_\R{a}^{p > 0}(t) \coloneqq \frac{1}{n_\R{p}} \sum_{i \in \C{I}_\R{p}} A_i^{p > 0}(t)$, and
$A_\R{a}^{p < 0}(t) \coloneqq \frac{1}{n_\R{p}} \sum_{i \in \C{I}_\R{p}} A_i^{p < 0}(t)$, with $A_i^{p > 0}$, $A_i^{p < 0}$ obtained for each participant $i$ through Algorithm \ref{alg:phase_estimation}.

\subsection{Results of the experiments}

\begin{figure}[t]
   \centering
   \includegraphics[scale=0.68]{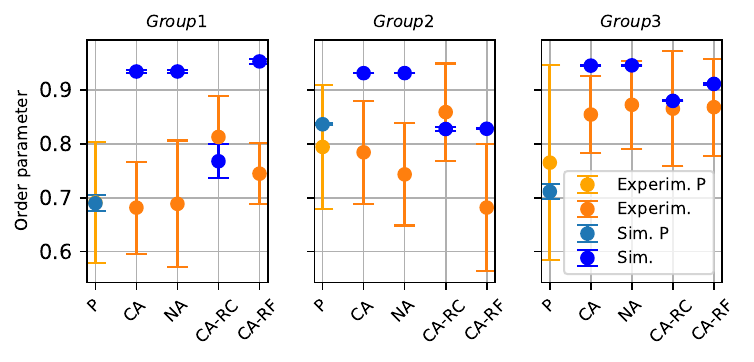}
   \caption{%
   Performance metrics for experiments in \S~\ref{sec:experiments}.
   We report $\langle r_{\R{net}}\rangle$ for conditions CA, NA, and $\langle r_{\R{tot}}\rangle$ for P, CA-RC, CA-RF.
   Gold and orange represent experimental data; blue and teal represent simulation data.
   }
   \label{fig:experiment_results}
\end{figure}

In Figure \ref{fig:experiment_results}, we present the performance metrics obtained under different experimental conditions (see the gold and orange lines). Although the cognitive architecture (CA) resulted in higher mean synchronization in some cases, such as in group 3 compared to the baseline condition P, a Wilcoxon test revealed no significant differences across most conditions relative to P, with the exception of CA-RC in group 1, where the difference was significant ($p = 0.03125$). 
This suggests the CA (and NA) typically neither disrupted nor improved synchronization significantly.

To more directly compare these results with those in Section~\ref{sec:simulations}, which demonstrated a consistent performance increase when using the CA, we estimated the coupling constant $c$ for each experimental group following the methodology in \cite{calabrese2021spontaneous}. 
Then, we ran simulations reproducing the experimental conditions: the 
results are depicted in Figure \ref{fig:experiment_results} (see the blue and teal lines).

In most scenarios, except for CA-RC and CA-RF in group 2, the application of the CA on synthetic data significantly enhanced synchronization, aligning with the findings from Section \ref{sec:simulations}. Additionally, a one-way ANOVA test indicated a significant difference between synthetic and real data ($p < 0.05$). After applying a Bonferroni correction, statistically significant differences were observed in the CA, NA, and CA-RF conditions across groups 1 and 2.

To evaluate participants' ability to recognize avatars, we conducted $z$-tests comparing their responses to hypothetical distributions derived from random guessing. The findings, detailed in Table \ref{tab:answers}, show that the CA was correctly identified more frequently than expected by chance only in the CA conditions for groups 1 and 3, and in the CA-RC condition for group 3, amounting to four cases out of nine.

  \begin{table}[t]
  \centering
  \begin{tabular}{l|llll}
      \toprule
      Group  & \multicolumn{4}{c}{\% correct answers ($p$-value)}\\
      {} & \emph{CA} & \emph{NA} & \emph{CA-RC} & \emph{CA-RF}\\
      \midrule
      1 & 44\%(0.015)  & 28\%(0.072) & 45\%(0.576) & 20\%(0.372) \\
      2 & 16\%(0.585) & 24\%(0.210) & 15\%(0.210) & 15\%(0.639)\\
      3 & 36\%(0.095) & 24\%(0.072) & 45\%(0.025) & 50\%(0.639)\\
      \bottomrule
  \end{tabular}
  \vspace{0.5em}
  \caption{%
  Answers to questionnaire to identify CA and NA in the experiments in \S~\ref{sec:experiments}.
  The baseline levels obtained by random guessing are 20\% for conditions CA, NA and 25\% for CA-RC, CA-RF.
  $p$-values computed with $z$-tests; null hypothesis: answers given randomly.  
  }
  \label{tab:answers}
  \end{table}
\section{Conclusions}

We presented a cognitive architecture (CA) to enhance synchronization in human-avatar groups performing periodic motor tasks; the CA is based on reinforcement learning
and is trained on synthetic data generated by Kuramoto models.
Comprehensive numerical simulations show the effectiveness of the CA in improving synchronization in a large variety of conditions.

Preliminary experiments with real participants revealed that the CA does not negatively affect synchronization but fails to produce a consistent improvement in all experimental situations, while generating motion that is almost indistinguishable from that of people.
These results suggest that while the Kuramoto model used to train the CA can capture dynamics such as the effect of interaction graphs on synchronization \citep{alderisio2017interaction}, it may not account for critical mechanisms necessary for effective avatar intervention, such as people's response to stimuli, limited attention capacity, and time-varying frequencies.



\bibliography{references}             

\appendix

\section{group synchronization 
index}

Let 
$\psi_{\R{tot}}(t) \coloneqq \arg q_{\R{tot}}(t)$ (see \S~\ref{sec:problem_description}) denote the average phase of the group.
Define 
$\phi_{\R{tot}, i}(t) \coloneqq \theta_i(t)-\psi_{\R{tot}}(t)$.
Let $T_\R{w} \in \BB{R}_{> 0}$ be length of a time-window, and, for $t \ge T_\R{w}$, define the time-average phasor
$s_{\R{tot}, i}(t) \coloneqq \frac{1}{T_\R{w}} \int_{t-T_\R{w}}^t e^{\imag \phi_{\R{tot}, i}(\tau)} \, \R{d}\tau$.
Compute the phase
$\bar{\phi}_{\R{tot}, i}(t) \coloneqq \arg s_{\R{tot}, i}(t)$,
which is the average distance of $\theta_i$ from the group, in the time-window $T_\R{w}$.
Let
$\Delta \phi_{\R{tot}, i}(t) \coloneqq \phi_{\R{tot}, i}(t) - \bar{\phi}_{\R{tot}, i}(t)$,
and let the \emph{total group synchronization index} \citep{calabrese2022modeling} be
\begin{equation}\label{eq:total_group_sync_index}
    \rho_{\R{tot}}(t) \coloneqq \frac{1}{n}\abs{ \sum_{i=1}^n e^{\imag \, \Delta \phi_{\R{tot},i}(t)}},
    \quad \forall t > T_\R{w};
\end{equation}
$\rho_{\R{tot}}(t) \approx 1$ (resp.~0) means participants oscillate with homogeneous (resp.~heterogeneous) angular speeds. 
The \emph{net synchronization index} is computed like the total one, but $q_\R{net}$ is used, rather than $q_\R{tot}$ and the summation in \eqref{eq:total_group_sync_index} is over $\C{I}_\R{p}$ rather then $\{1, \dots, n\}$.
Moreover, $\langle \rho_\R{net} \rangle \coloneqq \tfrac{1}{T-T_\R{w}} \int_{T_\R{w}}^T \rho_\R{net}(\tau) \, \R{d}\tau$, and
$\langle \rho_\R{tot} \rangle \coloneqq \tfrac{1}{T-T_\R{w}} \int_{T_\R{w}}^T \rho_\R{tot}(\tau) \, \R{d}\tau$.

\end{document}

%% file: support_files/preamble.tex
\usepackage[utf8]{inputenc}   
\usepackage[T1]{fontenc}      
\usepackage[svgnames]{xcolor} 
\usepackage{lipsum}	          
\usepackage{soul}             
\usepackage{natbib}           

\usepackage{amsfonts}   
\usepackage{mathtools}  
\usepackage{amssymb}   
\usepackage{bm}         
\usepackage{cases}     

\usepackage[caption=false,font=footnotesize]{subfig}
\usepackage{graphicx}	          
\usepackage{booktabs}   	      
\usepackage[export]{adjustbox}  
\usepackage[ruled,vlined,linesnumbered,algo2e]{algorithm2e}  
\usepackage{tikz}

\makeatletter
\let\old@ssect\@ssect 
\makeatother
\usepackage[hidelinks]{hyperref}  
\makeatletter
\def\@ssect#1#2#3#4#5#6{%
  \NR@gettitle{#6}
  \old@ssect{#1}{#2}{#3}{#4}{#5}{#6}
}
\makeatother

\graphicspath{{figures/}}         
\allowdisplaybreaks               

\makeatletter
\patchcmd{\@algocf@start}
  {-1.5em}
  {0pt}
  {}{}
\makeatother

\SetKwComment{tcc}{ {$\triangleright$} }{}  
\SetKwComment{tcp}{ {$\triangleright$} }{}  

\SetCommentSty{myCommentStyle}

\newcommand{\T}{^{\mathsf{T}}}
\newcommand{\R}[1]{\mathrm{#1}}
\newcommand{\ON}[1]{\operatorname{#1}}
\newcommand{\B}[1]{\if#1\relax\bm{#1}\else\mathbf{#1}\fi} 
\newcommand{\BB}[1]{\mathbb{#1}}
\newcommand{\C}[1]{\mathcal{#1}}
\newcommand{\abs}[1]{\left\lvert #1 \right\rvert}

\newcommand{\imag}{\mathsf{i}}